\documentclass[universe,article,accept,moreauthors]{Definitions/mdpi} 
\firstpage{1} 
\pubvolume{1}
\issuenum{1}
\articlenumber{0}
\pubyear{2026}
\copyrightyear{2026}
\datereceived{ } 
\daterevised{ } 
\dateaccepted{ } 
\datepublished{ } 

\Title{Extending the M33 Extinction Map with Radio and Far-Infrared Gas Observations}

\Author{Yuxi Wang $^{1}$*\orcidA{}}

\AuthorNames{Yuxi Wang}

\address{%
$^{1}$ \quad Department of Astronomy, College of Physics and Electronic Engineering, Qilu Normal University, Jinan 250200, China; yuxiwang@qlnu.edu.cn\\
$^{2}$ \quad Shandong Key Laboratory of Space Environment and Exploration Technology, China}

\corres{Correspondence: yuxiwang@qlnu.edu.cn}

\abstract{As the third-largest galaxy in the Local Group, M33 requires high-precision, wide-area extinction maps to recover the intrinsic luminosity and enable accurate studies, yet the available maps remain limited in coverage.
In this work, we extend the stellar-based extinction map within the Panchromatic Hubble Andromeda Treasury: Triangulum Extended Region (PHATTER) footprint to the outer disk by calibrating empirical $A_V$-H\,\textsc{i} and $A_V$-total hydrogen relations in the PHATTER overlap region.
Within the PHATTER footprint, we retain the stellar-based map at 6$^{\prime\prime}$ resolution, and outside the area covered by the stellar-based extinction map we produce an extended map at $\approx 12^{\prime\prime}$ resolution. 
The resulting map preserves a continuous morphology across the PHATTER boundary and reveals clear spiral-arm and interarm contrast and patchy dust substructures across a broader area. 
The median and maximum extinction values, $\mathrm{median}(A_V)\approx 0.91$ mag and $A_{V,\mathrm{max}} \approx 2.62$ mag, are consistent with the high-extinction range reported by previous M33 studies, while naturally reflecting differences in tracers and spatial coverage. 
The extinction map provides a practical, publicly available dataset for galaxy-wide extinction correction and for further studies in M33.}


\begin{document}



\section{Introduction}

Interstellar extinction is produced by the absorption and scattering of starlight by dust grains. 
It is a key observable for tracing dust composition and grain populations, and for establishing reliable corrections to photometric and spectroscopic measurements \citep{2003ARA&A..41..241D,2018ARA&A..56..673G}. 
Because dust directly affects brightness, colors, and inferred stellar parameters, extinction correction propagates into many derived quantities, including star formation rates, stellar masses, and age estimates \citep{2020ARA&A..58..529S}. 
For this reason, extinction maps are not only descriptive products of dust structure but also fundamental calibration tools for galaxy studies. 
For nearby galaxies, where individual stars can be resolved, spatially resolved extinction maps are especially important because they connect dust distributions to local environments such as spiral arms, interarm regions, and star-forming complexes \citep{2018ARA&A..56..673G,2026AJ....171..326W}.

In recent years, extinction mapping in nearby galaxies has developed rapidly. 
Advances in wide-field and high-resolution surveys have enabled stellar-based extinction measurements over increasingly large areas in Local Group galaxies (e.g., Magellanic Clouds \citep{2022MNRAS.511.1317C}; M31 \citep{2015ApJ...814....3D,2025AJ....169..250W}; M33 \citep{2026AJ....171..326W}). 
However, such stellar-based extinction maps are constrained by the sky coverage and depth of the tracers. 

In particular, M33 is the third largest galaxy in the Local Group after the Milky Way (MW) and M31 \citep{2000PASP..112..529V}, with a distance of about 840 kpc and a relatively favorable viewing geometry \citep{1973ugcg.book.....N,1991ApJ...372..455F}. 
The nuclear region of M33 has long been discussed as a nearby environment for studying starburst-like activity \citep{1999ApJ...519..165G}, while the disk provides a well-resolved example of a late-type spiral system \citep{1973ugcg.book.....N}. 
The study of extinction curves toward M33 have demonstrated non-negligible spatial variations, emphasizing that local dust properties are not uniform across the disk \citep{2022ApJS..260...41W}. 
The most recent extinction distribution, constructed with individually resolved red giant branch stars from the Panchromatic Hubble Andromeda Treasury: Triangulum Extended Region (PHATTER) survey \citep{2021ApJS..253...53W}, also reveals strongly inhomogeneous dust structures and reports positive spatial correspondence between extinction, gas distribution, and infrared emission \citep{2026AJ....171..326W}.
Despite the high spatial resolution, the map covers only the PHATTER footprint rather than the full disk, which could introduce uncertainties in galaxy-wide extinction correction, estimates of global dust properties, and comparisons between inner and outer interstellar environments.

In contrast, gas distributions (e.g., H\,\textsc{i} \citep{2010A&A...522A...3G} and total hydrogen \citep{2024A&A...688A.171K}) derived from far-infrared and radio observations provide much broader spatial coverage across M33.
To our knowledge, gas appears to be relatively co-located with the dust responsible for interstellar extinction \citep{2011piim.book.....D}, and the total hydrogen column scales with reddening as $\frac{N_{\mathrm H}}{E(B-V)}=5.8 \times 10^{21}\,{\mathrm H}\,\mathrm{cm}^{-2}\,\mathrm{mag}^{-1}$ in the MW \citep{1978ApJ...224..132B,2009ApJS..180..125R}.
Consequently, a positive correlation between extinction and the total hydrogen column density is expected on resolved galactic scales, while the normalization and scatter can vary with environment.
Moreover, on galactic scales dust and neutral gas are typically well mixed, with dust grains embedded in the gas \citep{2011piim.book.....D}.
Taken together, these points motivate us to adopt gas datasets to extend the stellar-based extinction map beyond the PHATTER footprint to the outer disk.

In this work, we construct an extended extinction map of M33 by combining the recent stellar-based extinction map within the PHATTER footprint derived by \citet{2026AJ....171..326W} with published wide-field H\,\textsc{i} \citep{2010A&A...522A...3G} and total hydrogen \citep{2024A&A...688A.171K} distributions, which were derived from Very Large Array (VLA) observations and Herschel flux maps with spatial resolutions of 12$^{\prime\prime}$ and 18.2$^{\prime\prime}$, respectively.
We calibrate the relations between extinction in the $V$ band, $A_V$, and gas column density in the overlap region, then adopt the calibrated trends to predict extinction outside the original stellar footprint, and finally merge the observed and predicted components into a single map. 
The resulting extinction map provides broader-area extinction information for M33 and enables more reliable photometric corrections and multi-wavelength studies throughout the galaxy.
We describe the data in Section \ref{sec:data} and detail the methodology in Section \ref{sec:method}. 
Section \ref{sec:results} presents the results and discussion.
Finally, the main conclusions are summarized in Section \ref{sec:conclusions}.

\section{Data} \label{sec:data}

In this work, we combine one stellar-based extinction map and two wide-field gas distributions to construct an extended extinction distribution with coverage out to the outer disk.

The adopted extinction map is the high-resolution M33 map constructed from resolved red giant branch stars in the PHATTER survey \citep{2021ApJS..253...53W} by fitting multiband color indexes in the F336W, F475W, F814W, F110W, and F160W bands \citep{2026AJ....171..326W}.
The color fitting employs extinction curves toward individual sight lines in M33 from \citet{2022ApJS..260...41W}, which allows spatial variations in the extinction law to be included in the extinction calculation.
The extinction map reaches an angular resolution of about $6^{\prime\prime}$, corresponding to $\sim 24$ pc at the distance of M33.
It resolves dust structures in spiral arms, interarm regions, and compact high-extinction clumps, and it has been shown to correlate positively with gas and infrared structures across the PHATTER-covered disk \citep{2026AJ....171..326W}.

To trace the large-scale atomic gas in M33, we adopt the H\,\textsc{i} map derived from VLA observations, constructed by integrating the H\,\textsc{i} spectral-line cube over the velocity range associated with M33 to obtain the projected intensity distribution of atomic gas across the disk \citep{2010A&A...522A...3G}. 
For this work, we use the product with a resolution of 12$^{\prime\prime}$, which corresponds to about 49 pc at the distance of M33.
Following standard 21 cm practice, the integrated intensity provided by the H\,\textsc{i} map, in units of $\mathrm{K\,km\,s^{-1}}$, is converted to column density via
\begin{equation}
N_{\mathrm{HI}}\,(\mathrm{cm^{-2}})=1.823\times10^{18}\,I_{\mathrm{HI}}\,(\mathrm{K\,km\,s^{-1}}),
\end{equation}
where $I_{\mathrm{HI}}$ is the velocity-integrated 21 cm brightness temperature.
The H\,\textsc{i} map provides a wide-field atomic-gas tracer that can be directly compared with extinction structures and applied to constrain extinction in the outer disk of M33.

To include the total gas component associated with dust emission, we use the M33 total hydrogen map with a resolution of 18.2$^{\prime\prime}$, derived by \citet{2024A&A...688A.171K} based on Herschel far-infrared observations.
The hydrogen column density, $N_{\mathrm{H}}$, in \citet{2024A&A...688A.171K} is inferred by performing dust spectral energy distribution (SED) fitting with 160, 250, 350, and 500 $\mu$m maps.
To reduce systematic biases in the gas reconstruction, spatial variations in the dust-emissivity parameters, including $\beta$ and $\kappa_0$, are explicitly accounted for \citep{2024A&A...688A.171K}.
The map complements the H\,\textsc{i} tracer by incorporating both atomic and molecular contributions in a dust-based framework, and it provides broad areal coverage suitable for extrapolating extinction beyond the PHATTER region.
Together with the H\,\textsc{i} map, it constitutes the gas data for calibrating and extending the extinction distribution on galaxy-wide scales.

\section{Method} \label{sec:method}

In this work, we extend the extinction map in M33 in two steps.
First, in the PHATTER overlap region, we calibrate extinction with gas tracers after matching the angular resolutions and map grids.
Second, assuming that the $A_V$-gas relations do not systematically differ between the inner and outer disk of M33, we apply the calibrated relations outside the PHATTER footprint to predict extinction across the outer region, and combine the two gas-based results on a common grid to construct a single, galaxy-wide extinction map.

\subsection{Calibration}\label{subsec:method_calib}

The stellar-tracer extinction map has a resolution of approximately $6^{\prime\prime}$ \citep{2026AJ....171..326W}, while the adopted H\,\textsc{i} and total hydrogen maps have resolutions of $12^{\prime\prime}$ \citep{2010A&A...522A...3G} and $18.2^{\prime\prime}$ \citep{2024A&A...688A.171K}, respectively.
We first convolve the $A_V$ map with Gaussian kernels to match the resolutions of H\,\textsc{i} and total hydrogen distributions, and reproject the smoothed extinction maps onto the H\,\textsc{i} and H grids.
Within the overlap region, we then calibrate two relations,
\begin{equation}
A^{\mathrm{(HI)}}_V=f_{\mathrm{HI}}(N_{\mathrm{HI}}),
\end{equation}
and
\begin{equation}
A^{\mathrm{(H)}}_V=f_{\mathrm{H}}(N_{\mathrm{H}}),
\end{equation}
with robust nonparametric trendlines instead of global linear fits.
The gas axis is sampled with sliding bins, and each bin is represented by the median $A_V$ after a 3$\sigma$ outlier rejection based on the median absolute deviation (MAD), where $\sigma = 1.4826 \times$ MAD and MAD = median($\lvert {A_V - \mathrm{median}(A_V)} \lvert $).
The cleaned nodes are then interpolated with a shape-preserving piecewise cubic Hermite interpolating polynomial (PCHIP; \citealt{1980SJNA...17..238F}). 
This approach preserves the overall monotonic tendency and avoids oscillatory artifacts that often arise with high-order polynomial fits.

\begin{figure}[H]
	\centering
	\subfloat{\includegraphics[width=0.48\textwidth]{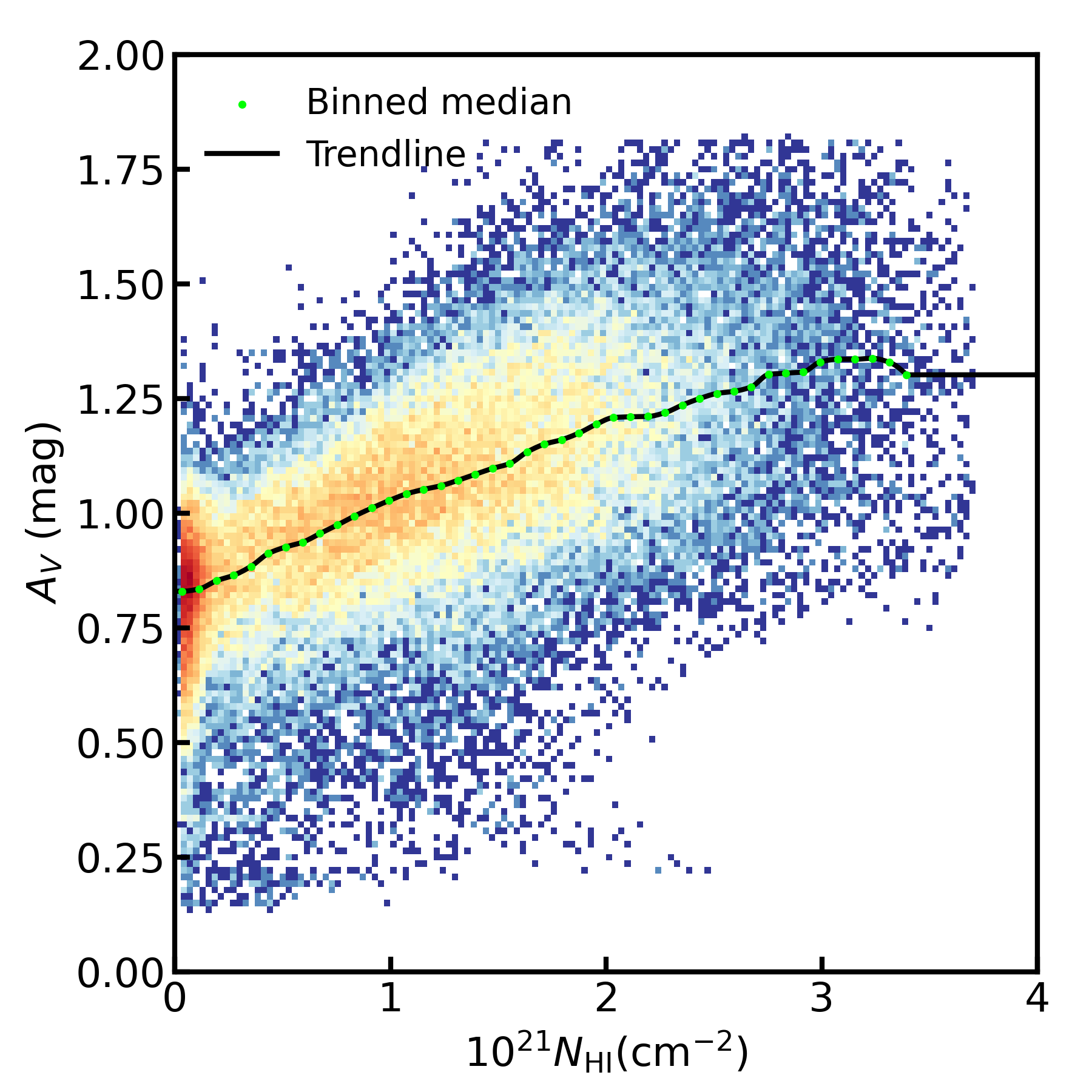}}\hfill
	\subfloat{\includegraphics[width=0.48\textwidth]{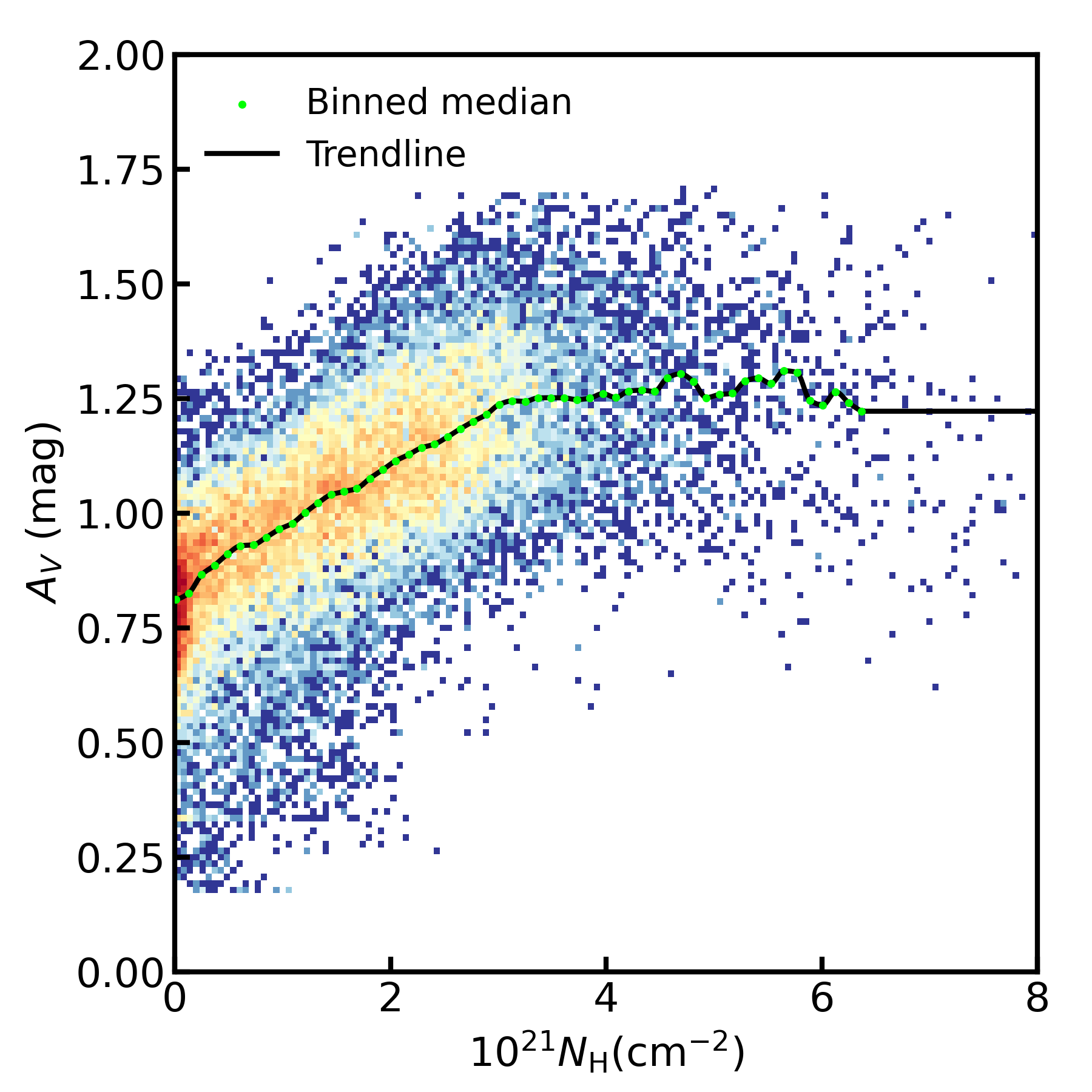}}
	\caption{Calibration of extinction with $N_{\mathrm{HI}}$ (left panel) and $N_{\mathrm{H}}$ (right panel) in the PHATTER region.
	The background colors show the two-dimensional pixel number density (redder indicates higher density). 
	Green points are robust binned medians after outlier rejection, and the black curves are the shape-preserving PCHIP trendlines adopted for the final calibration.} 
	\label{fig:method_calib}
	\end{figure}

The overlap regions contain 56,324 pixels for $A^{\mathrm{(HI)}}_V - N_{\mathrm{HI}}$ and 24,948 pixels for $A^{\mathrm{(H)}}_V - N_{\mathrm{H}}$ for the calibration.
The ranges of fitted nodes are $N_{\mathrm{HI},21}\in[0.0337,3.3937]$ and $N_{\mathrm{H},21}\in[0.0094,6.3694]$, in units of $10^{21}\,\mathrm{cm^{-2}}$.
Figure~\ref{fig:method_calib} presents the two calibration results, which show clear positive correlations with extinction, with the high-column regime departing from a single linear trend.


\subsection{Extinction Extension and Map Construction}\label{subsec:method_extend_build}

The calibrated functions are applied to the full gas maps to generate two predicted extinction maps, $A_V^{(\mathrm{HI})}$ and $A_V^{(\mathrm{H})}$, for regions outside the PHATTER footprint.
To avoid unconstrained extrapolation, gas values beyond the calibrated node ranges are clipped to the nearest boundary ($N^{\rm upper}_{\mathrm {HI}} = 3.3937 \times 10^{21}\,{\rm cm}^{-2}$, $N^{\rm lower}_{\mathrm {HI}} = 0.0337 \times 10^{21}\,{\rm cm}^{-2}$, $N^{\rm upper}_{\mathrm {H}} = 6.3694 \times 10^{21}\,{\rm cm}^{-2}$, $N^{\rm lower}_{\mathrm {H}} = 0.0094 \times 10^{21}\,{\rm cm}^{-2}$).
We then evaluate the calibrated PCHIP relation at the clipped value, yielding constant end behavior outside the calibrated range and preventing spurious extrapolation-driven trends.
The final extinction map is constructed on the H\,\textsc{i} grid, so the adopted effective resolution of the extended map is $12^{\prime\prime}$, corresponding to about 49 pc at the distance of M33.

On this common grid, the H\,\textsc{i}-based and H-based predictions are combined with inverse-variance weighting,
\begin{equation}
A_V^{\mathrm{pred}}=\frac{w_{\mathrm{HI}}A_V^{(\mathrm{HI})}+w_{\mathrm{H}}A_V^{(\mathrm{H})}}{w_{\mathrm{HI}}+w_{\mathrm{H}}},
\end{equation}
where
\begin{equation}
w_{\mathrm{HI}}=\sigma_{\mathrm{HI}}^{-2},
\qquad
w_{\mathrm{H}}=\sigma_{\mathrm{H}}^{-2}.
\end{equation}
The residual dispersions are measured over the PHATTER overlap by comparing each gas-based estimate with the observed extinction,
\begin{equation}
\sigma_{\mathrm{HI}}=\mathrm{std}\!\left(A_V^{\mathrm{obs}}-A_V^{(\mathrm{HI})}\right),
\qquad
\sigma_{\mathrm{H}}=\mathrm{std}\!\left(A_V^{\mathrm{obs}}-A_V^{(\mathrm{H})}\right),
\end{equation}
where “std” denotes the standard deviation over the overlap region.
In this work, we obtain $\sigma_{\mathrm{HI}} = 0.202138$ and $\sigma_{\mathrm{H}} = 0.199795$, indicating that the H\,\textsc{i}-based and H-based extinction estimates are weighted almost equally in the combination. 
Where only one gas tracer is observed, we use that single tracer's estimate.

Finally, we generate an observation-priority merged map.
Pixels within PHATTER keep the stellar-based extinction measurements, whereas pixels outside are filled with the weighted prediction defined above.
In this way, the inner disk extinction is determined by observations, while the map is physically extended to the outer disk with the calibrated gas relations.
It should be noted that the stellar-based extinction map constructed by \citet{2026AJ....171..326W} reflects the average extinction of the observed stars rather than the full line-of-sight dust column. 
Therefore, the extinction map constructed in this work provides the average extinction experienced by the stars in each pixel, instead of the full dust column along the sight line.
In practice, such a mean extinction map serves as a practical tool to provide extinction corrections for stellar observations and subsequent studies, especially in cases where the relative locations of the target stars and the dust along the line of sight are likewise unknown.

\section{Results and Discussion} \label{sec:results}
\subsection{Extended Extinction Distribution of M33}\label{subsec:results_map}

With the calibration and extension methodology described in Section~\ref{sec:method}, we derive a gas-extended extinction map of M33 and generate a single observation-priority merged map, expending the coverage by a factor of approximately 16.33 compared with the previous stellar-based extinction map constructed by \citet{2026AJ....171..326W}.
The final map is sampled on the H\,\textsc{i} grid and has an effective angular resolution of $12^{\prime\prime}$, corresponding to approximately 49 pc at the distance of M33.
Compared with the original PHATTER-limited extinction map, the extinction map derived in this work substantially increases the usable sky coverage and enables spatially continuous extinction correction from the inner disk to outer regions.

\begin{figure}[H]
	\centering
	\includegraphics[width=1\textwidth]{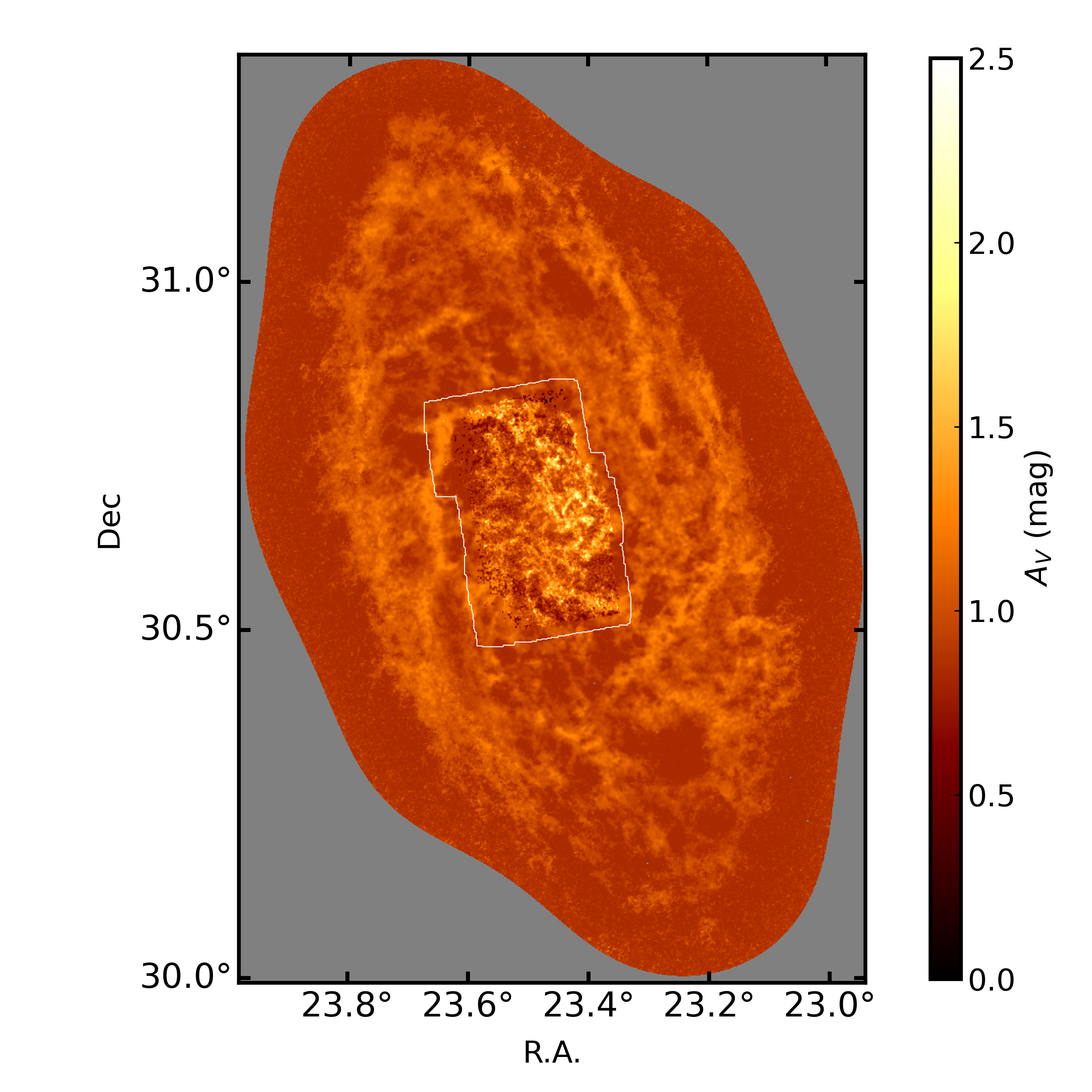}
	\caption{Extended extinction map of M33 with an effective angular resolution of \(12^{\prime\prime}\).
	The white boundary marks the PHATTER footprint.
	Within the PHATTER footprint, pixels adopt the stellar-based extinction measurements \citep{2026AJ....171..326W}.
	In the region not covered by the stellar-based extinction map, pixels are filled with the weighted, gas-calibrated prediction based on the H\,\textsc{i} and total hydrogen distributions \citep{2010A&A...522A...3G,2024A&A...688A.171K}.}
	\label{fig:result_ext_map}
	\end{figure}

Figure~\ref{fig:result_ext_map} presents the final extended extinction distribution, with a white outline enclosing the PHATTER footprint.
The map preserves the fine extinction structures measured from resolved stars within PHATTER and reveals coherent large-scale patterns outside the original stellar footprint. 
The morphology shows clear arm-interarm contrast and patchy dust structures over a wide area, indicating that the extension retains physically meaningful spatial variations rather than producing a smoothed and uniform background.
Moreover, the transition from the PHATTER region to the extended outer region is spatially continuous, with no obvious boundary artifacts introduced by the gas-based extension.

We provide two extinction maps on Zenodo (doi: \href{https://doi.org/10.5281/zenodo.22894226}{10.5281/zenodo.22894226}).
The first is the observation-priority merged map shown in Figure~\ref{fig:result_ext_map}.
The second is a prediction-only map that contains the gas-calibrated extension over the full available area.
For most applications, we recommend adopting the merged map as the default science product, because it preserves stellar-based \(A_V\) where available and relies on the gas-constrained extension only where stellar coverage is absent.
The Zenodo repository also includes a Python script that computes $V$-band extinction for specified coordinates (right ascension and declination), together with a usage example.

\subsection{Uncertainty Estimation} \label{subsec:uncertainty}

A quantitative uncertainty estimate is essential for assessing the reliability and scientific utility of the extended extinction map.
In this work, we calculate the uncertainty of the extinction distribution by propagating the scatter in the gas-extinction calibration relations.

\begin{figure}[H]
	\centering
	\includegraphics[width=0.82\textwidth]{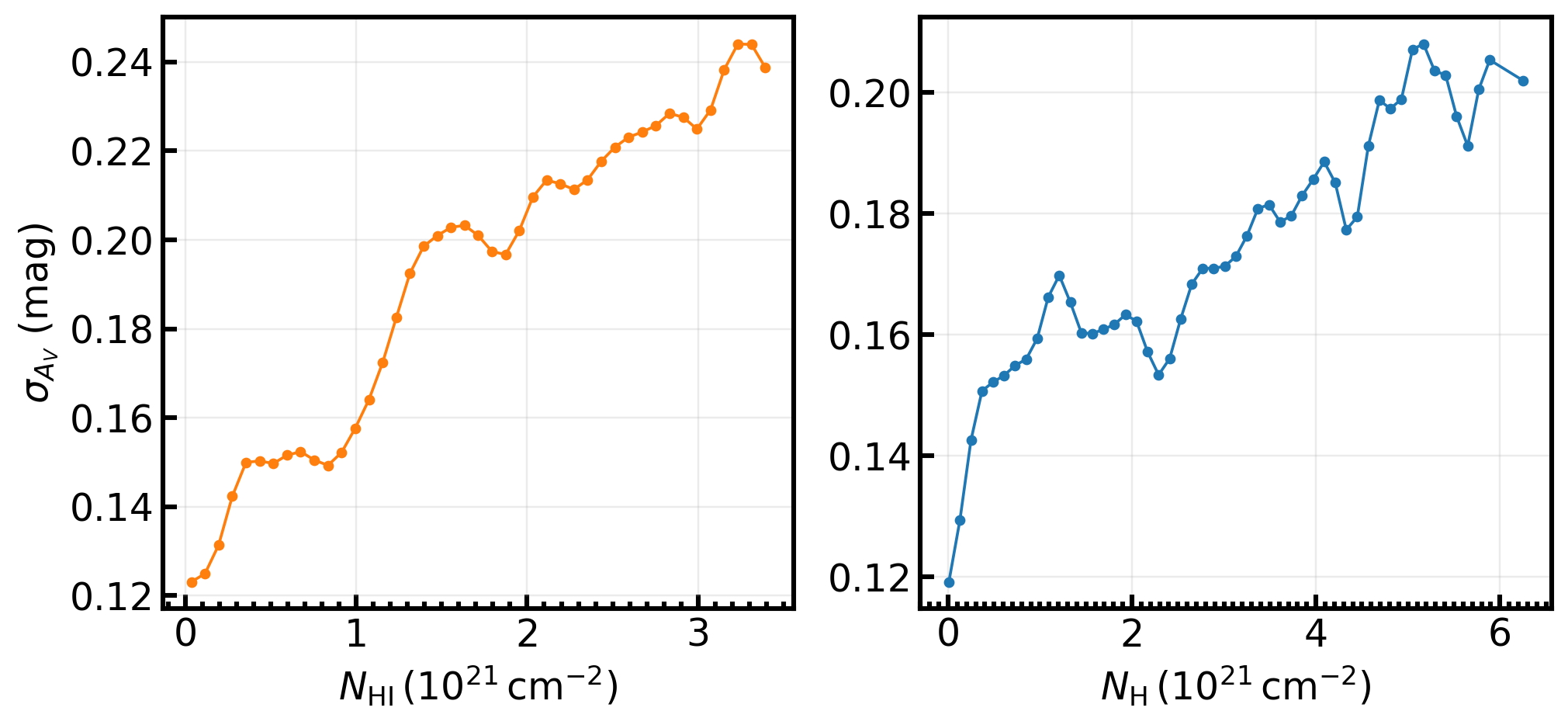}
	\caption{Calibration uncertainty of extinction as a function of $N_{\mathrm{HI}}$ (left panel) and $N_{\mathrm H}$ (right panel) in the PHATTER footprint.
	The points show local \(1\sigma\) residual measurements from sliding bins in the overlap region, and the curves are PCHIP interpolations following the same fitting strategy adopted in Section \ref{subsec:method_calib}.}
	\label{fig:calib_1sigma}
	\end{figure}

As mentioned in Section \ref{subsec:method_calib}, within the overlap between the extinction and gas maps, we derived two nonparametric calibration relations, \(A_V=f_{\mathrm{HI}}(N_{\mathrm{HI}})\) and \(A_V=f_{\mathrm{H}}(N_{\mathrm{H}})\).
In the same sliding gas bins used for calibration, in addition to taking the binned medians for the trends, we also quantify the dispersion with the scaled median absolute deviation (MAD) and define the local 1$\sigma$ ranges, $\sigma_{N_{\mathrm{HI}}}$ and $\sigma_{N_{\mathrm{H}}}$, shown as orange and blue points in Figure \ref{fig:calib_1sigma}, respectively.
The measurements are then converted into continuous functions of column density with the same shape-preserving PCHIP interpolation adopted in Section \ref{subsec:method_calib}, yielding the calibration uncertainties:
\begin{equation}
	\sigma^{(\mathrm{HI})}_{A_V}=f_{\sigma_{\mathrm{HI}}}(\sigma_{N_{\mathrm{HI}}}),
\end{equation}
and
\begin{equation}
	\sigma^{(\mathrm{H})}_{A_V}=f_{\sigma_{\mathrm{H}}}(\sigma_{N_{\mathrm{H}}}).
\end{equation}
As shown by the orange and blue curves of Figure \ref{fig:calib_1sigma}, the calibration uncertainties are not constant, but vary with gas column density.

The extinction uncertainty map is constructed with the same resolution as the extinction map in Section \ref{subsec:results_map}.
For each sky pixel, we evaluate the uncertainty function at the local gas column density.
Where both H\,\textsc{i} and total hydrogen measurements are available, we adopt an inverse-variance combination,
\begin{equation}
	\sigma_{A_V} = \sqrt{(\frac{w_{\mathrm{HI}}}{w_{\mathrm{HI}}+w_{\mathrm{H}}})^2 (\sigma^{\mathrm{HI}}_{A_V})^2 + (\frac{w_{\mathrm{H}}}{w_{\mathrm{HI}}+w_{\mathrm{H}}})^2 (\sigma^{\mathrm{H}}_{A_V})^2},
\end{equation}
where $w_{\mathrm{HI}}=\sigma_{\mathrm{HI}}^{-2}$ and $w_{\mathrm{H}}=\sigma_{\mathrm{H}}^{-2}$ as mentioned in Section \ref{subsec:method_extend_build}.
In pixels with only one gas tracer, we take the corresponding uncertainty directly.

\begin{figure}[H]
\centering
\includegraphics[width=0.68\textwidth]{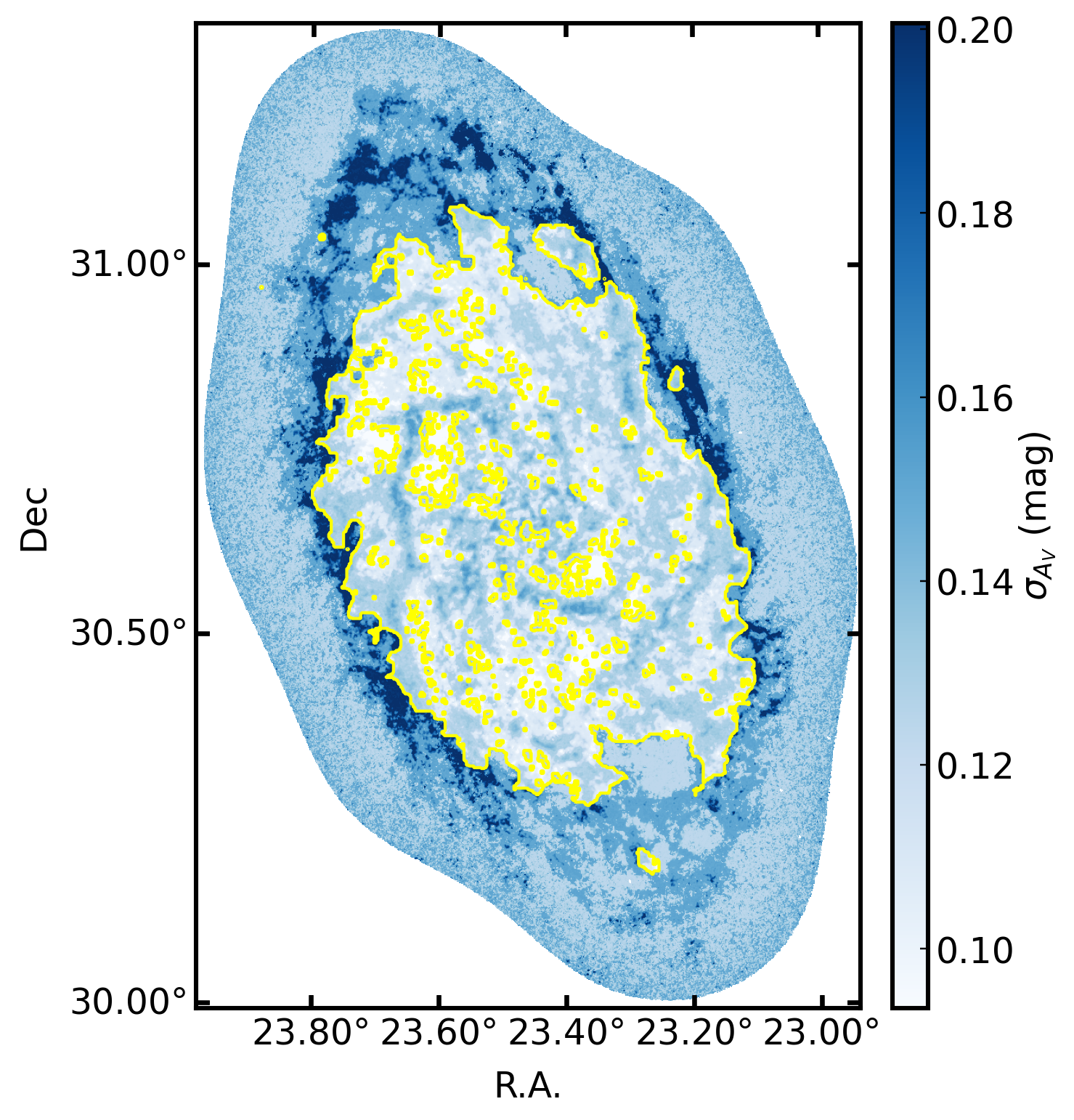}
\caption{Spatial distribution of uncertainty for the extinction predicted from the calibration relations.
The yellow contour encloses pixels where both H\,\textsc{i} and total hydrogen observations are available.
Uncertainties are generally lower inside the contour and higher outside where only one gas tracer is available.}
\label{fig:uncert_1sigma_map}
\end{figure}

Figure~\ref{fig:uncert_1sigma_map} presents the resulting uncertainty distribution.
We emphasize that the map quantifies the uncertainty of the extinction predicted from the calibration relations, not the uncertainty of the merged map that prioritizes observations shown in Figure \ref{fig:result_ext_map}. 
The yellow contour outlines the area where both H\,\textsc{i} and total H data are available, inside which the uncertainties are systematically smaller, whereas outside they are generally larger in regions constrained by a single gas tracer.

The distribution of extinction uncertainties is also provided on Zenodo as referenced in Section \ref{subsec:results_map}. 
The script included on Zenodo can likewise be applied to compute the extinction uncertainty at specified positions.

\subsection{Quantitative Comparison with Previous Works} \label{subsec:com}

Quantitatively, we obtain a median extinction of 0.91 mag\endnote{
We analyze the histogram of the extinction map constructed in this work and find that the distribution of $A_V$ is not strictly lognormal, unlike results reported for other galaxies \citep{2009AJ....137.3000H, 2023AJ....165..166K, 2025AJ....170..195R}.
However, based on the Akaike Information Criterion (AIC), the $A_V$ distribution is better described by a lognormal than by a Gaussian. 
Accordingly, we adopt the median rather than the mean of $A_V$.} 
and a maximum value of $A_{V,\mathrm{max}} = 2.62$ mag for the extinction map derived in this work.
For comparison, the extinction map in the PHATTER footprint constructed by \citet{2026AJ....171..326W} reports $\langle A_V\rangle \approx 1.05$ mag and $A_{V,\mathrm{max}} \approx 2.5$ mag.
The extended map constructed in this work therefore preserves the high-extinction tail while yielding a slightly lower global mean, as expected: the outer disk is dominated by low-to-moderate extinction, whereas the highest-extinction clumps remain concentrated in the inner disk and along spiral arms.

We also compare the extinction map derived in this work with a sequence of previous M33 studies that adopted different tracers and methods.
With multiwavelength attenuation measurements in radial bins, \citet{verley2009} found a relatively low average of $A_V \approx 0.25$ mag. 
In contrast, moving to stellar-based extinction, \citet{2017PhDT.......221H} reported a higher mean of $\approx 0.53$ mag with a maximum of about 2.5 mag.
Focusing on young clusters, \citet{2022AJ....163...16M} showed that two-thirds have $E(B-V) < 0.2$ mag, indicating generally low to moderate extinction for young, luminous populations.
Complementarily, modeling OB-type supergiants from UV to NIR, \citet{2022ApJS..260...41W} obtained a mean $A_V \approx 0.43$ mag while still reaching an upper end of $\approx 2.5$ mag. 
Table~\ref{tab:av_comparison_literature} summarizes these studies alongside this work.

\begin{table}[H]
	\centering
	\caption{Comparison of characteristic M33 extinction values and methods between this work and representative previous studies.}
	\label{tab:av_comparison_literature}
	\begin{tabular}{p{3cm}p{4.0cm}p{2.8cm}p{2.8cm}}
	\hline
	Study & Data and method & Mean/Median extinction & Maximum extinction \\
	\hline
	This work (full extended map) & PHATTER-based map extended with H\,\textsc{i} and H calibration & median$(A_V)$ = 0.91 mag & $A_{V,\mathrm{max}} = 2.62$ mag \\
	Wang et al. (2026) \citep{2026AJ....171..326W} & HST PHATTER resolved-star photometry with RGB-tracer fitting & $\langle A_V\rangle \approx 1.05$ mag & $A_{V,\mathrm{max}} \approx 2.5$ mag \\
	Wang et al. (2022) \citep{2022ApJS..260...41W} & OB-type supergiants with photometry from UV to NIR, extinction fitting toward individual sight lines & $\langle A_V\rangle \approx 0.43$ mag & $A_{V,\mathrm{max}} \approx 2.5$ mag \\
	Moeller \& Calzetti (2022) \citep{2022AJ....163...16M} & Young star clusters with UV-IR photometry and SED-based extinction estimates & two-thirds with $E(B-V)<0.2$ mag & $E(B-V) \approx 0.6$ mag \\
	Hagen (2017) \citep{2017PhDT.......221H} & Stellar-sample-based extinction analysis in M33 & $\langle A_V\rangle \approx 0.53$ mag & $A_{V,\mathrm{max}} \approx 2.5$ mag \\
	Verley et al. (2009) \citep{verley2009} & Radial-bin attenuation from H$\alpha$, UV, and IR emission & $\langle A_V\rangle \approx 0.25$ mag & not reported \\
	\hline
	\end{tabular}
	\end{table}

As explained in \citet{2026AJ....171..326W}, a crucial factor behind the differing mean/median extinctions is the choice of tracer. 
In this work, we extend the stellar-based extinction map derived by \citet{2026AJ....171..326W}, which adopted RGB stars as tracers.
RGB stars are brighter than lower main-sequence stars, which makes them detectable through higher extinction, and they belong to an older population with a larger vertical scale height than young clusters or OB supergiants \citep{2013A&ARv..21...61R,2016ARA&A..54..529B}, allowing us to probe deeper dust columns along the sight line.
As a result, the average extinction values derived in this work and in \citet{2026AJ....171..326W} are higher than those derived from younger tracers\endnote{Based on the PHATTER observations, we also select main-sequence stars and helium-burning stars and construct an extinction map for the PHATTER footprint in M33 following the same method as \citet{2026AJ....171..326W}. 
The resulting mean extinction is 0.44 mag, which is consistent with the mean values reported by \citet{2022ApJS..260...41W} adopting OB supergiants and \citet{2022AJ....163...16M} with young clusters.
Therefore, under comparable observational conditions, we infer that the mean extinction obtained with RGB stars is roughly two to three times that derived from young tracers.}.
Moreover, RGB stars are broadly distributed across the disk, which makes the resulting spatially averaged extinction more representative at the galaxy scale.
High spatial resolution and the large sample of individually resolved RGB stars and gas tracers also help recover heavily obscured regions that integrated or lower-resolution approaches may smooth out\endnote{
We additionally adopt the member stars derived by \citet{2021ApJ...907...18R} to construct an extinction distribution in M33. 
The catalog of member stars is dominated by late-type stars, and it is based on observations from the United Kingdom Infra-Red Telescope, which is a ground-based facility. 
The spatial resolution is therefore much lower than that of space-based imaging. 
The resulting map covers about 1 deg$^2$, and the mean extinction is approximately 0.44 mag, which is about one half of the mean value obtained in this work, indicating that the spatial resolution of the tracer data can also bias the inferred mean.}. 
It is worth noting that the resulting maximum $A_V$ is consistent with the upper values reported by \citet{2022ApJS..260...41W} and \citet{2017PhDT.......221H}, supporting that the map constructed in this work captures the broad dynamic range of extinction in M33.

\section{Conclusions} \label{sec:conclusions}

Building on the stellar-based PHATTER extinction map and wide-field H\,\textsc{i} and total hydrogen data, we construct a galaxy-wide extinction map of M33. 
Inside the PHATTER footprint, we directly adopt the extinction map of \citet{2026AJ....171..326W} at the native $\approx 6^{\prime\prime}$ resolution. 
In the overlap between PHATTER and the gas data, we calibrate the $A_V$-gas relations and apply them beyond the coverage of the stellar-based extinction map  to produce an extended map with an effective resolution of $\approx 12^{\prime\prime}$. 
The morphology presents clear arm-interarm contrast and patchy dust substructure over a much wider disk area, and remains continuous across the original boundary.

The extinction map constructed in this work has a median extinction of $ A_V \approx 0.91$ mag and a maximum of $A_{V,\mathrm{max}}\approx 2.62$ mag, preserving the high-extinction upper end of the stellar-based extinction map within the PHATTER footprint, while yielding a slightly lower global mean because the outer disk contributes more moderate extinction.
Compared with earlier work based on young populations or integrated attenuation, it remains consistent at the high-extinction end and helps explain differences in mean/median extinction driven by tracer choice and coverage. 
In practice, the extended map supports galaxy-wide extinction corrections, and serves as a practical dataset for further studies such as SED fitting and star formation measurements across M33.

\vspace{6pt}

\funding{This research was funded by the National Natural Science Foundation of China, grant number 12303030, the Shandong Provincial Natural Science Foundation, grant number ZR2024QA236,  the National Natural Science Foundation of China, grant number 12133002, and Youth Innovation Team  Science and Technology Support Project in Colleges and Universities of  Shandong Province, grant number 2025KJG043.}

\dataavailability{All input datasets adopted in this work are publicly available in the literature (stellar-based extinction map of M33: \citealt{2026AJ....171..326W}; H\,\textsc{i} map: \citealt{2010A&A...522A...3G}; total hydrogen map: \citealt{2024A&A...688A.171K}). 
The M33 extinction map constructed in this study, together with the uncertainty map, is available on Zenodo (doi: \href{https://doi.org/10.5281/zenodo.22894226}{10.5281/zenodo.22894226}).}

\acknowledgments{
We are very grateful to the anonymous reviewers for the constructive comments and suggestions, which have significantly improved this work.
It is a pleasure to thank Profs. Jian Gao and Bingqiu Chen for the invaluable discussions.}

\conflictsofinterest{The authors declare no conflicts of interest. 
The funders had no role in the design of the study; in the collection, analyses, or interpretation of data; in the writing of the manuscript; or in the decision to publish the results}

\begin{adjustwidth}{-\extralength}{0cm}
\printendnotes[custom] 

\reftitle{References}



\bibliography{references}

%


\PublishersNote{}
\end{adjustwidth}
\end{document}